\documentclass{aa}
\usepackage{amsmath}
\usepackage[varg]{txfonts}
\usepackage{graphicx,amssymb,float,subfigure}
\usepackage{natbib}
\bibpunct{(}{)}{;}{a}{}{,}
\begin{document}

\title{On the applicability of the level set method beyond the
       flamelet regime in thermonuclear supernova simulations}

\author{W. Schmidt}

\institute{Lehrstuhl f\"{u}r Astronomie, Institut f\"{u}r Theoretische Physik und Astrophysik,
  Universit\"{a}t W\"{u}rzburg, Am Hubland, D-97074 W\"{u}rzburg, Germany}

\date{Received / Accepted}

\titlerunning{On the applicability of the level set method beyond the
       flamelet regime}

\abstract{In thermonuclear supernovae, intermediate mass elements
are mostly produced by distributed burning provided that a
deflagration to detonation transition does not set in. Apart from
the two-dimensional study by \citet{RoepHille05}, very little
attention has been payed so far to the correct treatment of this
burning regime in numerical simulations. In this article, the
physics of distributed burning is reviewed from the literature on
terrestrial combustion and differences which arise from the very
small Prandtl numbers encountered in degenerate matter are pointed
out. Then it is shown that the level set method continues to be
applicable beyond the flamelet regime as long as the width of the
flame brush does not become smaller than the numerical cutoff
length. Implementing this constraint with a simple parameterisation
of the effect of turbulence onto the energy generation rate, the
production of intermediate mass elements increases substantially
compared to previous simulations, in which the burning process was
stopped once the mass density dropped below
$10^{7}\,\mathrm{g\,cm^{-3}}$. Although these results depend on the
chosen numerical resolution, an improvement of the constraints on the
the total mass of burning products in the pure deflagration scenario
can be achieved.

\keywords{Stars: supernovae: general --
Hydrodynamics -- Turbulence -- Methods: numerical} }

\maketitle

\section{Introduction}

In the course of the last few years, observational indications in
favour of a delayed detonation in type Ia supernovae (SNe Ia) have
mounted. For example, calculations of the X-ray spectrum of the
Tycho supernova remnant assuming various hydrodynamical models
appear to support a deflagration to detonation transition (DDT)
\citep{BadBor06}. Furthermore, an investigation of near-infrared
emission lines of three branch-normal supernovae by
\citet{MarHoef06} implies very little carbon residuals at radial
velocities less than $1.8\cdot 10^{4}\,\mathrm{cm\,s^{-1}}$. Even
the most advanced three-dimensional simulations of thermonuclear
supernovae assuming pure deflagrations \citep{RoepHille06,SchmNie06}
fail to satisfy this constraint. In models with delayed detonations
\citep{GamKhokh06}, on the other hand, the supersonic propagation of
burning fronts dispose of virtually all carbon except for the
outermost layers. Furthermore, a gravitational confined detonation was
suggested as an alternative scenario \citep{PlewCald04}.

However, a recent numerical study by \cite{MaierNie06} demonstrated
that detonation waves fail to penetrate processed material stemming
from the initial deflagration phase. Therefore, pockets of unburned
material are likely to survive even a delayed detonation. Apart from
that, \citet{niemeyer1999a} pointed out several theoretical
arguments against DDTs. In addition, accommodating the observed
variability of SNe Ia \citep{StriLeib06} within the DDT scenario
appears to be difficult, because delayed detonations tend to produce
at least a solar mass of iron group elements and explosion energies
in excess of $10^{51}\,\mathrm{erg}$, which are typical values
characterising bright SNe Ia. But there are also SNe Ia of moderate
or low luminosity producing much smaller masses of iron-group
elements and less explosion energy. Apart from that,
\citet{JhaBranch06} inferred non-negligible amounts of carbon at low
expansion velocities from the late-time spectroscopy of the SN
2002cx, a peculiar supernova with very low luminosity. If this
result was confirmed by further observations, one might conceive of
a sub-class of SNe Ia originating either from Chandrasekhar mass
white dwarfs which are only partially burned by pure deflagrations
or from sub-Chandrasekhar mass progenitors. In this article, we
shall be concerned exclusively with the Chandrasekhar mass scenario.

\citet{SchmNie06} showed that the explosion energy and mass of iron
group elements in thermonuclear supernova simulations with pure
deflagration can be varied over about an order of magnitude if
non-simultaneous point ignitions are applied. To that end, the
simple MLT model of the pre-supernova core proposed by
\citet{WunWoos04} was adopted for the implementation of a stochastic
ignition procedure. In a numerical case study, models with a total
number of ignitions ranging from a few up to several hundred events
per octant were investigated. The main result is that the total mass
of iron group elements can be adjusted to any value smaller than
$0.75 M_{\sun}$ with a maximal explosion energy of roughly
$0.8\cdot10^{51}\,\mathrm{erg}$. Therefore, these models are
feasible candidates for less energetic SNe Ia.

Due to the artificially chosen termination of thermonuclear burning
at mass densities below $10^{7}\,\mathrm{g\,cm^{-3}}$, however, the
prediction of the carbon and oxygen residuals in the models with
stochastic ignition by \citet{SchmNie06} was not reliable. The
transition from the flamelet regime to the regime of distributed
burning is expected to occur at a mass density of about $3\cdot
10^{7}\,\mathrm{g\,cm^{-3}}$ \citep{NieWoos97,NieKer97}. Since the
level set method--as implemented by \citet{ReinHille99a}--was
applied for the numerical flame front propagation, the treatment of
the distributed burning regime remained unclear. As a first
approximation, distributed burning was mostly suppressed by
introducing the aforementioned density threshold. This resulted in
an overestimate of the amount of unburned material. Apart from that,
a delayed detonation might be triggered in the late burning phase
\citep{NieWoos97,KhokOran97}. However, the arguments put forward by
\citet{niemeyer1999a} and the numerical results by \citet{BellDay04}
shed serious doubt on this proposition. For this reason, it appears
even more important to consider the possibility of the subsonic
distributed burning mode at lower densities.

A first attempt to include distributed burning in deflagration
models of SNe Ia was made by \citet{RoepHille05}. They carried out
two-dimensional simulations demonstrating that the explosion energy
increases and the fraction of carbon and oxygen at low radial
velocities is reduced. However, their prescription of the turbulent
burning speed in the distributed regime suffered from a
misconception of the relevant scales. In this article, we will argue
that the existing treatment of burning fronts with the level set
method can be carried over to the distributed regime provided that
the burning time scale is smaller than the eddy turn-over time scale
associated with the numerical cutoff length. Consequently, this
constraint has to be observed in numerical simulations of
thermonuclear supernovae with the level set method.

After reviewing the physics of distributed burning in
Section~\ref{sc:distrb_burn}, we will show in
Section~\ref{sc:ext_level_set} that the turbulent burning speed
continues to be determined by the magnitude of turbulent velocity
fluctuations at unresolved scales after the onset of distributed
burning. In Section~\ref{sc:num_simul}, we will discuss results from
numerical simulations of thermonuclear supernovae with stochastic
ignition, where the burning process was terminated based on a
comparison of the burning and unresolved eddy turn-over time
scales. Thereby, it was possible to increase the production of
intermediate mass elements significantly compared to previous
simulations with a density threshold of $10^{7}\,\mathrm{g\,cm^{-3}}$
for thermonuclear burning.  In particular, we found that the total
mass of intermediate mass elements appears to be independent of the
details of the ignition process, although slower ignition implies a
larger amount of fuel at the transition from the flamelet to the
distributed burning regime.

\section{Distributed burning}
\label{sc:distrb_burn}

In combustion physics, two different regimes of deflagration are
distinguished: On the one hand, the \emph{flamelet regime}, in which
the microscopic flame propagation speed is solely determined by the
thermal conductivity of the fuel. In the \emph{distributed burning
regime}, on the other hand, the transport of heat and mass is
influenced by turbulence even at scales comparable to the width of
the flame brush. As argued by \citet{NieKer97}, the correct
criterion for the transition from burning in the flamelet regime to
distributed burning is that the flame width $\delta$ is about the
Gibson length $\ell_{\mathrm{G}}$. The Gibson length is implicitly
defined by the equality of the velocity $v'(\ell)$ associated with
turbulent eddies of size $\ell$ and the laminar flame speed
$s_{\mathrm{lam}}$  \citep{Peters88}:
\begin{equation}
    \label{eq:gibson}
    v'(\ell_{\mathrm{G}})=s_{\mathrm{lam}}.
\end{equation}
In the flamelet regime, the flame front propagates sufficiently fast
through eddies smaller than the Gibson length such that turbulence
has virtually no effect on the internal structure of the reaction
zone. The width of the flame is then given by
\begin{equation}
    \label{eq:flame_width}
    \delta=C_{\delta}\sqrt{\chi\tau_{\mathrm{nuc}}},
\end{equation}
where $\chi$ is the thermal conductivity, $C_{\delta}$ a constant
dimensionless coefficient and the thermonuclear reaction time scale
is defined by
\begin{equation}
    \label{eq:time_nuc}
    \tau_{\mathrm{nuc}} = \frac{\rho\epsilon}{Q},
\end{equation}
Here $\epsilon$ is the specific thermonuclear energy release and $Q$
the rate of energy generation per unit time and unit volume. The
time scale $\tau_{\mathrm{nuc}}$ and the conductivity $\chi$ also
determine the laminar flame speed:
\begin{equation}
    \label{eq:lam_speed_cond}
    s_{\mathrm{lam}} = \sqrt{\frac{\chi}{\tau_{\mathrm{nuc}}}}.
\end{equation}

Once $l_{\mathrm{G}}\sim\delta$, however, turbulence will begin to
affect the transport of heat due to turbulent mixing of preheated
material into fuel. As a consequence of the enhanced diffusivity,
the flame front is broadened. If the turbulence intensity does not
become too high, however, only the preheat zone will be broadened
while the reaction zone is not disturbed by turbulent eddies due to
the increased viscosity near the flame. This was, for instance,
demonstrated in numerical simulations by \citet{SanMen00}. The
effect of turbulence onto the flame can be quantified in terms of a
turbulent diffusivity $\chi^{\ast}$ and the width $\delta^{\ast}$ of
the broadened preheating zone. According to \emph{Damk\"{o}hler's
hypothesis}, the propagation speed $s_{\mathrm{lam}}^{\ast}$ of the
broadened flame is enhanced in proportion to the square root of the
turbulent diffusivity relative to the thermal diffusivity
\citep{Dam40}:
\begin{equation}
    \label{eq:speed_lam_enh}
    \frac{s_{\mathrm{lam}}^{\ast}}{s_{\mathrm{lam}}}=
    \left(\frac{\chi^{\ast}}{\chi}\right)^{1/2}.
\end{equation}
This relation follows exactly if the rate of energy generation $Q$
is unaffected by turbulence. Since the width of the reaction zone
tends to become significantly smaller than $\delta^{\ast}$ (by a
factor greater than 10), this mode of burning is called the
\emph{thin-reaction-zones regime}.

The conductivity $\chi$ can be related to the viscosity $\nu$ by
$\chi=\mathrm{Pr}\nu$, where the \emph{Prandtl number} is a
dimensionless characteristic number of the fuel. In stark contrast
to most terrestrial fluids which have a Prandtl number of the
roughly unity, $\mathrm{Pr}\ll 1$ for degenerate carbon and oxygen
in white dwarfs \citep{NandPe84}. Introducing the turbulent
viscosity $\nu^{\ast}=\nu_{\mathrm{t}}+\nu$, the turbulent
diffusivity can be expressed as
$\chi^{\ast}=(\nu_{\mathrm{t}}+\nu)/\mathrm{Pr^{\ast}}$, where
$\nu_{\mathrm{t}}$ accounts for the effect of turbulent eddies in
the range of length scales from the Kolmogorov scale
$\eta_{\mathrm{K}}$ to the $\delta^{\ast}$. Thus,
equation~(\ref{eq:speed_lam_enh}) can be written in the form
\begin{equation}
    \label{eq:speed_lam_enh_visc}
    \frac{s_{\mathrm{lam}}^{\ast}}{s_{\mathrm{lam}}} =
    \left(\frac{\mathrm{Pr}}{\mathrm{Pr}^{\ast}}\right)^{1/2}
    \left(\frac{\nu_{\mathrm{t}}}{\nu}+1\right)^{1/2}.
\end{equation}

Assuming that turbulence becomes asymptotically isotropic towards
scales small compared to the integral scale, we may use
$\mathrm{Pr}^{\ast}\sim 1$ for the turbulent Prandtl number
\citep{YakOrs86} and the velocity fluctuations $v'(\ell)$ at
sufficiently small length scales $\ell$ asymptotically obey the
Kolmogorov-Obukhov 1/3-law \citep{Frisch},
\begin{equation}
    \label{eq:one_third_law}
    v'(\ell)\sim(\epsilon\ell)^{1/3}.
\end{equation}
Numerical evidence for asymptotic isotropy and the above power law
in simulations of Rayleigh-Taylor-unstable thermonuclear flames was reported
by \citet{ZingWoos05}.
The rate of energy dissipation $\epsilon=\nu^{3}/\eta_{\mathrm{K}}$,
where the Kolmogorov scale $\eta_{\mathrm{K}}$ is about the size of
the smallest eddies. Thus, substitution of
$(\epsilon\ell_{\mathrm{G}})^{1/3}$ on the left hand side of
equation~(\ref{eq:gibson}) yields
\begin{equation}
    \label{eq:gibson_relation}
    \ell_{\mathrm{G}}^{1/3}\delta\sim\frac{1}{\mathrm{Pr}}\eta_{\mathrm{K}}^{4/3}.
\end{equation}
Since $\ell_{\mathrm{G}}\sim\delta$ upon the onset of distributed
burning, it follows from the above relation that
$\delta^{\ast}\gtrsim\delta\sim\mathrm{Pr}^{-3/4}\eta_{\mathrm{K}}$.
In thermonuclear supernovae, $10^{-5}\lesssim\mathrm{Pr}\lesssim
0.1$ \citep{NandPe84}. Thus, the the broadened flame is much wider
than the Kolmogorov scale.

Assuming approximate local equilibrium of the energy flux through
the cascade of turbulent eddies from size $\delta^{\ast}$ down to
the Kolmogorov scale, we have
$\nu_{\mathrm{t}}/\nu\simeq(\delta^{\ast}/\eta_{\mathrm{K}})^{4/3}\gg
1$. Equation~(\ref{eq:speed_lam_enh_visc}) for the enhanced flame
propagation speed then implies
\begin{equation}
    \label{eq:speed_lam_enh_kolmgrv}
    \frac{s_{\mathrm{lam}}^{\ast}}{s_{\mathrm{lam}}}\sim
    \left(\frac{\mathrm{Pr}}{\mathrm{Pr}^{\ast}}\right)^{1/2}
    \left(\frac{\delta^{\ast}}{\eta_{\mathrm{K}}}\right)^{2/3}.
\end{equation}
Note that the asymptote $s_{\mathrm{lam}}^{\ast}\simeq
s_{\mathrm{lam}}$ corresponds to
$\delta^{\ast}\simeq\mathrm{Pr}^{-3/4}\eta_{\mathrm{K}}$ which is
consistent with the estimate in the previous paragraph. Since the
definition of the flame width~(\ref{eq:flame_width}) implies
$\chi^{1/2}\propto\delta$, it follows form
Damk\"{o}hler's relation~(\ref{eq:speed_lam_enh}) that
$s_{\mathrm{lam}}^{\ast}/s_{\mathrm{lam}}\sim \delta^{\ast}/\delta$.
Combining this result with
relation~(\ref{eq:speed_lam_enh_kolmgrv}), it is possible to
eliminate the width of the broadened flame $\delta^{\ast}$ and
calculate $s_{\mathrm{lam}}^{\ast}$ as a function of known
parameters. For a detailed specification of the coefficients and the
resulting expression for $s_{\mathrm{lam}}^{\ast}$ we refer to
\citet{KimMen00}.

In terms of time scales, the thin-reaction-zones regime is
characterised by
\begin{equation}
    \tau_{\mathrm{nuc}}\sim\frac{\ell_{\mathrm{G}}}{v'(\ell_{\mathrm{G}})}\sim
    \frac{\eta_\mathrm{K}^{2}}{\mathrm{Pr}^{1/2}\nu},
\end{equation}
where the expression on the very right follows from the
1/3-law~(\ref{eq:one_third_law}) and
relation~(\ref{eq:gibson_relation}) with
$\delta\sim\ell_{\mathrm{G}}$. For constant viscosity, growing
energy flux corresponds to a decreasing Kolmogorov scale. However,
as the time scale $\mathrm{Pr}^{-1/2}\eta_\mathrm{K}^{2}/\nu$
becomes much smaller than $\tau_{\mathrm{nuc}}$, turbulent eddies
will penetrate and eventually break the reaction zone. In this
\emph{broken-reaction-zones regime}, turbulence entrains preheated
fuel with patches of already burned material at a time scale smaller
than the nuclear time scale~(\ref{eq:time_nuc}). It appears that the
dilution of burning material within the reaction zone would imply a
modification of the rate of energy generation per unit volume.  As a
simple parametrisation, we set $Q^{\ast}=C^{\ast}Q$, where $Q$ is
the corresponding energy generation rate if turbulence was absent.
In the case of premixed combustion, which evidently applies to
thermonuclear fusion, the entrainment of fuel with ash may inhibit
but cannot boost the burning process, i.e. $C^{\ast}\lesssim 1$. In
any case, the quantitative determination of $C^{\ast}$ requires
three-dimensional direct numerical simulations of turbulent burning
with a detailed reaction network at different densities and varying
turbulence intensity. Numerical investigations pointing in this
direction were presented by \citet{BellDay04}. In conclusion,
writing the effective reaction time scale as
$\tau^{\ast}=\tau_{\mathrm{nuc}}/C^{\ast}$, we have
\begin{equation}
    \label{eq:time_scl_broken_react}
    \tau^{\ast}\gtrsim\tau_{\mathrm{nuc}}\gg
    \frac{\eta_\mathrm{K}^{2}}{\mathrm{Pr}^{1/2}\nu}
\end{equation}
in the broken-reaction-zones regime.

\section{Extended level set prescription}
\label{sc:ext_level_set}

\citet{ReinHille99b} introduced the level set method for the
modelling of burning fronts in thermonuclear supernova simulations.
The basic idea of this approach is to represent the flames by the
zero level set of a distance function that is determined by a
partial differential equation. The intrinsic propagation speed in
the flamelet regime is the laminar burning speed $s_{\mathrm{lam}}$.
Because simulations of SNe Ia are essentially large eddy
simulations, however, the effective propagation speed is
asymptotically proportional to the magnitude of unresolved turbulent
velocity fluctuations, i.e. $v'(\Delta)$, where $\Delta$ is the
resolution of the numerical grid \citep{NieHille95}. For this
reason, the effective propagation speed of the numerically computed
flame fronts is called the turbulent flame speed and denoted by
$s_{\mathrm{t}}$. For the calculation of $s_{\mathrm{t}}$, we have
adopted Pocheau's model \citep{Poch94,SchmNie06b}:
\begin{equation}
  \label{eq:flame_speed_poch}
  s_{\mathrm{t}} = s_{\mathrm{lam}}
  \sqrt{1 + C_{\mathrm{t}}
        \left(\frac{q_{\mathrm{sgs}}}{s_{\mathrm{lam}}}\right)^{2}},
\end{equation}
For fully developed turbulence, the flame propagation speed is
asymptotically given by
$s_{\mathrm{t}}\simeq\sqrt{C_{\mathrm{t}}}q_{\mathrm{sgs}}$. LES of
turbulent combustion in a box indicated that a sound burning rate
was obtained with $C_{\mathrm{t}}$ not differing too far from unity
\citep{SchmHille05}. We adopted the value $C_{\mathrm{t}}=4/3$ which
is consistent with \citet{Peters99}. The subgrid scale turbulent
velocity $q_{\mathrm{sgs}}\sim v'(\Delta)$ is computed by means of
the localised subgrid scale model proposed by \citep{SchmNie06a}. As
a great benefit for the application to thermonuclear supernova
simulations, this subgrid scale model does not rely on certain flow
properties such as isotropy or a particular mechanism for the
production of turbulence on large scales.

According to \citet{KimMen00}, the above model for the turbulent
flame speed can be readily extended into the thin-reaction-zones
regime:
\begin{equation}
  s_{\mathrm{t}}^{\ast} = s_{\mathrm{lam}}^{\ast}
  \sqrt{1 + C_{\mathrm{t}}
        \left(\frac{q_{\mathrm{sgs}}^{\ast}}{s_{\mathrm{lam}}^{\ast}}\right)^{2}}.
\end{equation}
Here, the laminar flame speed is substituted by the enhanced flame
propagation speed $s_{\mathrm{lam}}^{\ast}$ introduced in
Section~(\ref{sc:distrb_burn}). Since $s_{\mathrm{lam}}^{\ast}$
incorporates the effects of turbulence at length scales
$\ell\lesssim\delta^{\ast}$, the contribution of turbulent velocity
fluctuations in this range of length scales has to be subtracted
from $q_{\mathrm{sgs}}$. This results in the reduced subgrid scale
turbulence velocity $q_{\mathrm{sgs}}^{\ast}$ corresponding to the
range of length scales between the width of the broadened flame,
$\delta^{\ast}$, and the numerical resolution. In the case of
Kolmogorov scaling~(\ref{eq:one_third_law}),
\begin{equation}
    \frac{q_{\mathrm{sgs}}^{\ast}}{q_{\mathrm{sgs}}}\simeq
    \left[1-\left(\frac{\delta^{\ast}}{\Delta}\right)^{2/3}\right]^{1/2}.
\end{equation}

The transition from the flamelet to the distributed burning regime
is expected to occur within the range of densities from $10^{7}$ to
$10^{8}\,\mathrm{g/cm^{3}}$. The calculations of \citet{TimWoos92}
show that $\delta$ assumes values between $10^{-2}$ and a few
$\mathrm{cm}$ within this range of densities. The width
$\delta^{\ast}$ of the broadened flame cannot exceed $\delta$ by
more than a factor $\sim 10$ without turbulent eddies breaking up
the reaction zone \citep{KimMen00}. Thus, it appears that
$\delta^{\ast}\lesssim 10^{2}\,\mathrm{cm}$ for the
thin-reaction-zones regime in thermonuclear supernovae. Even in the
most elaborate numerical simulations, on the other hand,
$\Delta\gtrsim 10^{5}\,\mathrm{cm}$. Thus, $q_{\mathrm{sgs}}^{\ast}$
differs from $q_{\mathrm{sgs}}$ by a few percent at most and the
broadened flame may still very well be represented by a sharp
discontinuity. Moreover, the enhanced flame speed is small compared
to $q_{\mathrm{sgs}}^{\ast}\simeq q_{\mathrm{sgs}}$, because
$s_{\mathrm{lam}}^{\ast}/s_{\mathrm{lam}}=\delta^{\ast}/\delta\lesssim
10$ and $s_{\mathrm{lam}}\sim 1\,\mathrm{km/s}$ for $\rho\sim 10^{7}
\,\mathrm{g/cm^{3}}$ \citep{TimWoos92}, whereas
$q_{\mathrm{sgs}}\sim 100\,\mathrm{km/s}$ \citep{SchmNie06b}.
Therefore, the flame speed model~(\ref{eq:flame_speed_poch}) can be
carried over unaltered to the thin-reaction-zones regime with an
accuracy better than about 10 \%, which is within the intrinsic
uncertainties of the model.

As the exploding white dwarf expands further and the density drops
below $\sim10^{7} \,\mathrm{g/cm^{3}}$, the nuclear time scale rises
rapidly and, inevitably, the
condition~(\ref{eq:time_scl_broken_react}) for the
broken-reaction-zone regimes will be met. The broadened flame will
then dissolve into a flame brush in which fuel and ash are mixed by
turbulent eddies. Although there is no well defined width
$\delta^{\ast}$ of the flame brush, we can specify a turbulent
mixing length $\ell_{\mathrm{burn}}$ which corresponds to the
typical size of eddies with turn-over time of the order of the
effective reaction time $\tau^{\ast}$:
\begin{equation}
    \frac{\ell_{\mathrm{burn}}}{v'(\ell_{\mathrm{burn}})}
    \sim\tau^{\ast}.
\end{equation}
The length scale $\ell_{\mathrm{burn}}$ can be regarded as
generalisation of the Gibson length.

Since the width of the flame in the thin-reaction-zones regime is
very small compared to typical grid resolutions in simulations of
thermonuclear supernovae, the flame brush in the
broken-reaction-zones regime will initially not be resolved either,
i.e. $\ell_{\mathrm{burn}}\lesssim\Delta$. In terms of time scales,
$\tau^{\ast}=\tau_{\mathrm{nuc}}/C^{\ast}\lesssim\Delta/q_{\mathrm{sgs}}$.
As long as this constraint is satisfied, it appears to be a sensible
approximation to propagate the unresolved turbulent flame brush with
the level set method. For
\begin{equation}
    \label{eq:terminate}
    \tau_{\mathrm{nuc}}\gtrsim\frac{C^{\ast}\Delta}{q_{\mathrm{sgs}}},
\end{equation}
on the other hand, the level set representation breaks down. Thus,
without implementing an explicit treatment of volume-burning in the
code, the burning process has to be terminated in a numerical
simulation once the above condition is met. Of course, this implies
a dependence of the termination of the burning process on the
numerical resolution. Without including volume-burning, however,
this implication is inevitable. Former numerical simulations of
thermonuclear supernovae with the level set method merely appeared
to be converged because of an artifical termination criterion that
was chosen to be independent of the resolution (see the following
Section).

The extension of the level set method into the distributed burning
regime as outlined in this Section markedly differs from the flame
speed model applied by \citet{RoepHille05}, which is based on
Damk\"{o}hler's hypothesis~(\ref{eq:speed_lam_enh}) and,
consequently, is restricted to the thin-reaction-zones regime.
Essentially, their model is based on the assumption
$s_{\mathrm{t}}\simeq s_{\mathrm{lam}}^{\ast}$ or, equivalently,
$\delta^{\ast}\simeq\Delta$. It should become clear from the above
discussion, however, that this assumption does not hold.

\begin{table}[htb]
  \caption{Power-law fit of the nuclear energy timescale $\tau_{\mathrm{nuc}}$ in the
    range $0.01\le\rho_{9}\le 0.5$. The numerical data are taken from
    \citet{TimWoos92}.}
  \label{tb:fit}
  \begin{center}
    \begin{tabular}{c c c}
      \hline
      $\rho_{9}$ & $\tau_{\mathrm{nuc}}$ (numerical) & $\tau_{\mathrm{nuc}}$ (fitted)\\
      \hline\hline
      0.5 & 0.00089 & 0.00126 \\
      0.2 & $8.66\cdot 10^{-6}$ & $3.06\cdot 10^{-6}$ \\
      0.1 & $1.18\cdot 10^{-7}$ & $2.29\cdot 10^{-7}$ \\
      0.05 & $1.76\cdot 10^{-8}$ & $1.71\cdot 10^{-8}$ \\
      0.01 & $5.23\cdot 10^{-10}$& $5.55\cdot 10^{-10}$ \\
      \hline
    \end{tabular}
  \end{center}
\end{table}

\begin{figure*}[thb]
  \begin{center}
    \includegraphics{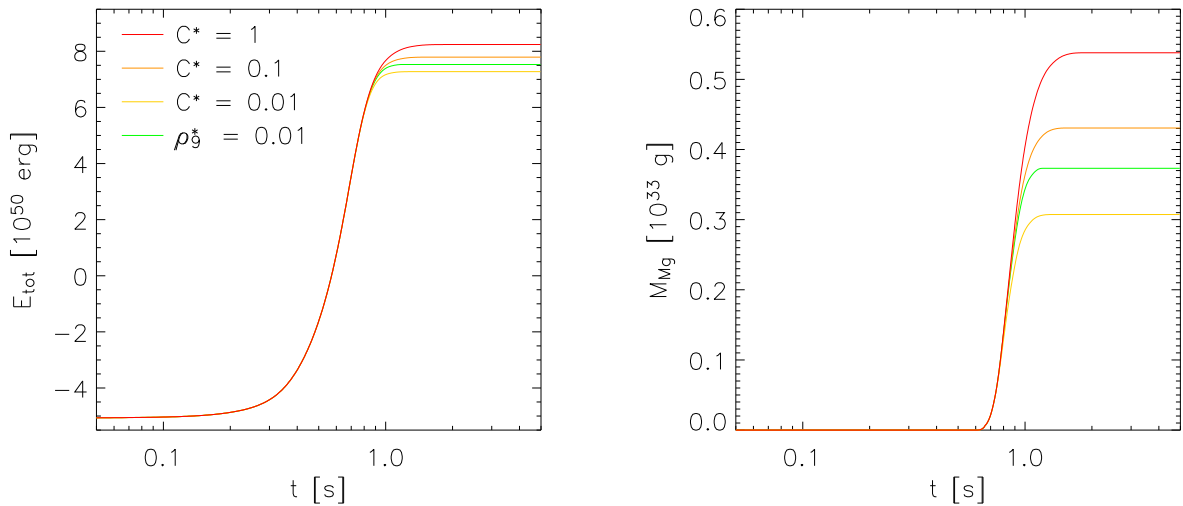}
    \caption{Single-octant simulations with $N=128^{3}$, $C_{\mathrm{e}}=10^{4}$
        and various choices of $C^{\ast}$. Plotted are the
        evolution of the total energy and the integrated
        mass of intermediate mass elements, respectively. Also
        shown is a reference simulation, in which the burning
        process was stopped at the density threshold $\rho_{9}\le
        0.01$.}
    \label{fg:evol_oct_m4}
  \end{center}
\end{figure*}

\begin{table*}[htb]
  \caption{Total energy release and masses of burning products at $t=5.0\,\mathrm{s}$
    for the simulations shown in Fig.~\ref{fg:evol_oct_m4}}.
  \label{tb:oct_m4}
  \begin{center}
    \begin{tabular}{l c c c c c}
      \hline
      & $E_{\mathrm{nuc}}\,[10^{51}\,\mathrm{erg}]$ & $E_{\mathrm{kin}}\,[10^{51}\,\mathrm{erg}]$ &
      $M_{\mathrm{Ni}}/M_{\sun}$ & $M_{\mathrm{Mg}}/M_{\sun}$ \\
      \hline\hline
      $\rho_{9}^{\ast}=0.01$ & 1.261 & 0.753 & 0.707 & 0.188 \\
      $C^{\ast} = 0.01$      & 1.235 & 0.727 & 0.708 & 0.155 \\
      $C^{\ast} = 0.1$       & 1.287 & 0.779 & 0.708 & 0.217 \\
      $C^{\ast} = 1.0$       & 1.332 & 0.824 & 0.708 & 0.270 \\
      \hline
    \end{tabular}
  \end{center}
\end{table*}

\section{Numerical simulations}
\label{sc:num_simul}

We investigated the effect of including burning with the extended
level set prescription in a follow-up study of \citet{SchmNie06}. To
that end, we modified the criterion for the termination of the
burning process in the code as follows. From the values of
$s_{\mathrm{lam}}$ and $\delta$ calculated by \citet{TimWoos92}, we
obtained $\tau_{\mathrm{nuc}}=\delta/s_{\mathrm{lam}}$ and linearly
fitted $\log\tau_{\mathrm{nuc}}$ as a function of $\log\rho_{9}$ ,
where $\rho_{9}$ is the mass density in units of
$10^{9}\,\mathrm{g/cm^{3}}$, for $\rho_{9}<1.0$. The power law
resulting from this fit is
\begin{equation}
    \label{eq:tau_nuc_fit}
    \tau_{\mathrm{nuc}}(\rho_{9}) =
    \frac{4.146\cdot 10^{-11}\,\mathrm{s}}{\rho_{9}^{3.7417}}.
\end{equation}
The data points as well as the corresponding values of the fit
function are listed in Table~\ref{tb:fit}. Evaluating
$\tau_{\mathrm{nuc}}(\rho_{9})$ locally in each cell, it can be
checked whether the criterion~\ref{eq:terminate} is fulfilled for a
certain prescribed constant $C^{\mathrm{\ast}}$. If the nuclear time
scale becomes larger than the threshold given by the right-hand
side, burning will be terminated. Since $\Delta\sim
10^{6}\,\mathrm{cm}$ and $q_{\mathrm{sgs}}\sim
10^{7}\,\mathrm{cm\,s^{-1}}$ after about one second,
$C^{\ast}\Delta/ q_{\mathrm{sgs}}\lesssim 0.1\,\mathrm{s}$. Thus,
the power law~\ref{eq:tau_nuc_fit} has to be somewhat extrapolated
towards $\rho_{9}\approx 0.003$, which appears to be reasonable.

To begin with, we computed a series of single-octant simulations on
grids consisting of $N=128^{3}$ cells with the code described in
\citet{SchmNie06b}. Due to the hybrid geometry of these grids, with a
fine-resolved uniform inner part and exponentially growing cell size
in the outer part, in combination with the co-expanding grid technique
introduced by \citep{Roep05}, it was possible to study trends even
with a relatively small number of grid cells. If we had chosen a
higher resolution, the applicability of the level set method would
have been constrained even further. This can be seen from the
criterion~(\ref{eq:terminate}) with the scaling law
$q_{\mathrm{sgs}}\sim \Delta^{1/3}$. Since the nuclear time scale
$\tau_{\mathrm{nuc}}$ gradually increases as the explosion progresses,
lowering the cutoff length $\Delta$ implies that the termination point
will be met earlier. On the other hand, a certain resolution is
mandatory for capturing the large-scale production of turbulence by
Rayleigh-Taylor instabilities. From the resolution study in
\citet{SchmNie06b} and the influence of doubling the resolution of
thermonuclear supernova simulations with stochastic ignition discussed
by \citet{SchmNie06}, we conclude that $128^{3}$ cells per octant are
sufficient for arriving at a sensible estimate of the total amount of
burning products.

\begin{figure*}[thb]
  \begin{center}
    \includegraphics{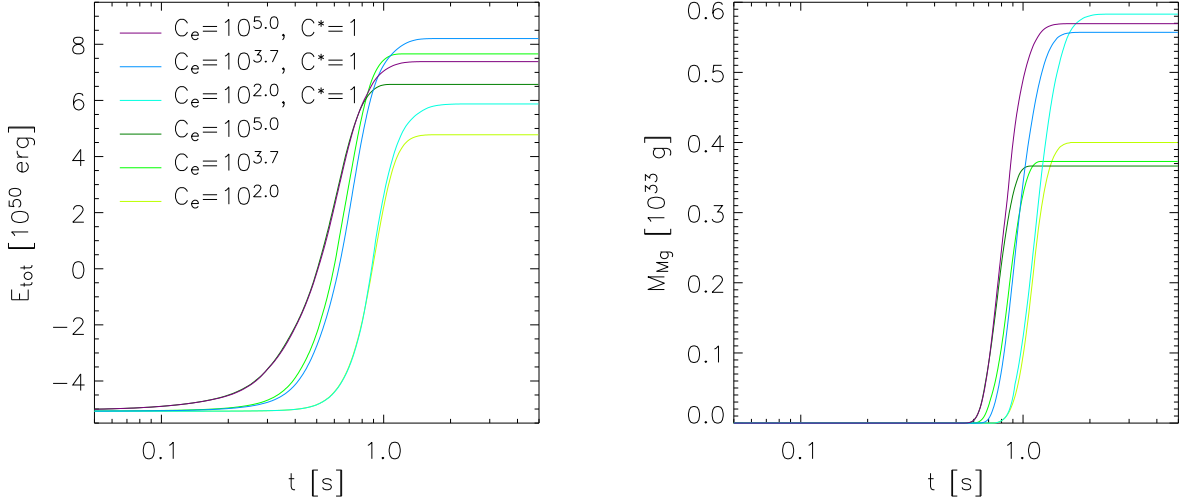}
    \caption{Evolution of the total energy and the integrated
        mass of intermediate mass elements, respectively, for
        several full-star simulations with $N=256^{3}$. For
        each value of the exponentiation parameter
        $C_{\mathrm{e}}$, which controls the number of
        ignitions events, either $\rho_{9}\le 0.01$ or
        $\tau_{\mathrm{nuc}}\ge\Delta/q_{\mathrm{sgs}}$
        (corresponding to $C^{\ast}=1.0$) was
        set for the termination of the burning process.}
    \label{fg:evol_4pi}
  \end{center}
\end{figure*}

\begin{table*}[htb]
  \caption{Total energy release and masses of burning products at $t=5.0\,\mathrm{s}$
    for the simulations shown in Fig.~\ref{fg:evol_4pi}}.
  \label{tb:evol_4pi}
  \begin{center}
    \begin{tabular}{l r c c c c c}
      \hline
      & $C_{\mathrm{e}}$ &
      $E_{\mathrm{nuc}}\,[10^{51}\,\mathrm{erg}]$ & $E_{\mathrm{kin}}\,[10^{51}\,\mathrm{erg}]$ &
      $M_{\mathrm{Ni}}/M_{\sun}$ & $M_{\mathrm{Mg}}/M_{\sun}$ \\
      \hline\hline
      $\rho_{9}^{\ast}=0.01$ & $10^{2}$        & 0.985 & 0.478 & 0.523 & 0.201 \\
      $\rho_{9}^{\ast}=0.01$ & $5\cdot 10^{3}$ & 1.274 & 0.766 & 0.715 & 0.188 \\
      $\rho_{9}^{\ast}=0.01$ & $10^{5}$        & 1.165 & 0.657 & 0.647 & 0.184 \\
      $C^{\ast} = 1.0$       & $10^{2}$        & 1.095 & 0.587 & 0.545 & 0.293 \\
      $C^{\ast} = 1.0$       & $5\cdot 10^{3}$ & 1.328 & 0.821 & 0.701 & 0.280 \\
      $C^{\ast} = 1.0$       & $10^{5}$        & 1.246 & 0.738 & 0.645 & 0.286 \\
      \hline
    \end{tabular}
  \end{center}
\end{table*}

Applying the stochastic ignition procedure formulated by
\citet{SchmNie06}, we selected a reference model with
$C_{\mathrm{e}}=10^{4}$, where the parameter $C_{\mathrm{e}}$
adjusts the time scale of Poisson processes generating the ignition
events in thin adjacent spherical shells. Depending on the choice of
$C_{\mathrm{e}}$, the outcome of the explosion, in particular, the
release of nuclear energy and the total mass of iron group elements
is varying greatly. For $C_{\mathrm{e}}=10^{4}$, the total number of
ignition events per octant is roughly $100$ and the yield of nuclear
energy becomes nearly maximal (see Table~1 and Fig.~1 in
\citet{SchmNie06}). We repeated this simulation with the same
initial white dwarf model and parameter settings, except for the
termination of the burning process as explained above, using three
different values of $C^{\ast}$.

The time evolution of the total amount of intermediate mass elements
(represented by ${^{24}}$Mg) plotted in the right panel of
Fig.~\ref{fg:evol_oct_m4} clearly shows the influence of the burning
termination criterion. Assuming $C^{\ast}=1$, the total magnesium
mass $M_{\mathrm{Mg}}$ at $t=5.0\,\mathrm{s}$ increases by $44\,\%$
compared to the reference simulation (see Table~\ref{tb:oct_m4}, in
which burning ceases for $\rho_{9}\le\rho_{9}^{\ast}=0.01$. One
should note that this is only a lower bound, because $C^{\ast}=1$
indicates that the broken-reaction-zones regime has just been
entered and volume burning at resolved scales, which cannot be
treated with the present code, would consume even more carbon and
oxygen. The mass of iron group elements (${^{54}}$Ni), on the other
hand, remains unaffected which, of course, reflects nickel being
produced entirely within the flamelet regime at densities
$\rho_{9}\ge 0.05$ while magnesium is stemming largely from
distributed burning at lower densities. Even for $C^{\ast}=0.1$,
meaning that the extended level set prescription breaks down well
within the broken-reaction-zones regime, the yield of intermediate
mass elements is still higher than in the case of the density
threshold $\rho_{9}^{\ast}=0.01$. If $C^{\ast}=0.01$, on the other
hand, $M_{\mathrm{Mg}}$ becomes slightly smaller. Since $C^{\ast}\ll
1$ implies a greatly reduced energy generation rate, the physical
quenching of the flame brush would be nearly reached in this case
and, as a consequence, only little volume burning could possibly
follow. In this case, we can be fairly sure that the actual amount
of burning products would be more or less what is seen in the
simulation.

We also performed full-star simulations with $N=256^{3}$, for which
$\Delta$ is initially the same as in the single-octant simulations.
Varying the exponentiation parameter $C_{\mathrm{e}}$, the burning
process was terminated once
$\tau_{\mathrm{nuc}}\ge\Delta/q_{\mathrm{sgs}}$, i.e.
$C^{\ast}=1.0$. The results in comparison to the corresponding
simulations with the termination criterion $\rho_{9}\le 0.01$ from
\citep{SchmNie06} are plotted in Fig.~\ref{fg:evol_4pi}. Because of
the substantially larger mass of unburned carbon and oxygen that
will be left over at the transition from iron group to intermediate
mass element production if $C^{\ast}$ is smaller and, consequently,
the ignition proceeds slower, one would expect $M_{\mathrm{Mg}}$ to
increase with $C_{\mathrm{e}}$. However, the burning statistics
listed in Table~\ref{tb:evol_4pi} clearly demonstrates that
$M_{\mathrm{Mg}}$ is almost constant for varying $C_{\mathrm{e}}$.
Hence, distributed burning appears to consume about the same amount
of fuel independent of the state of the exploding star at the end of
the flamelet regime.

\begin{figure}[thb]
  \begin{center}
    \resizebox{\hsize}{!}{\includegraphics{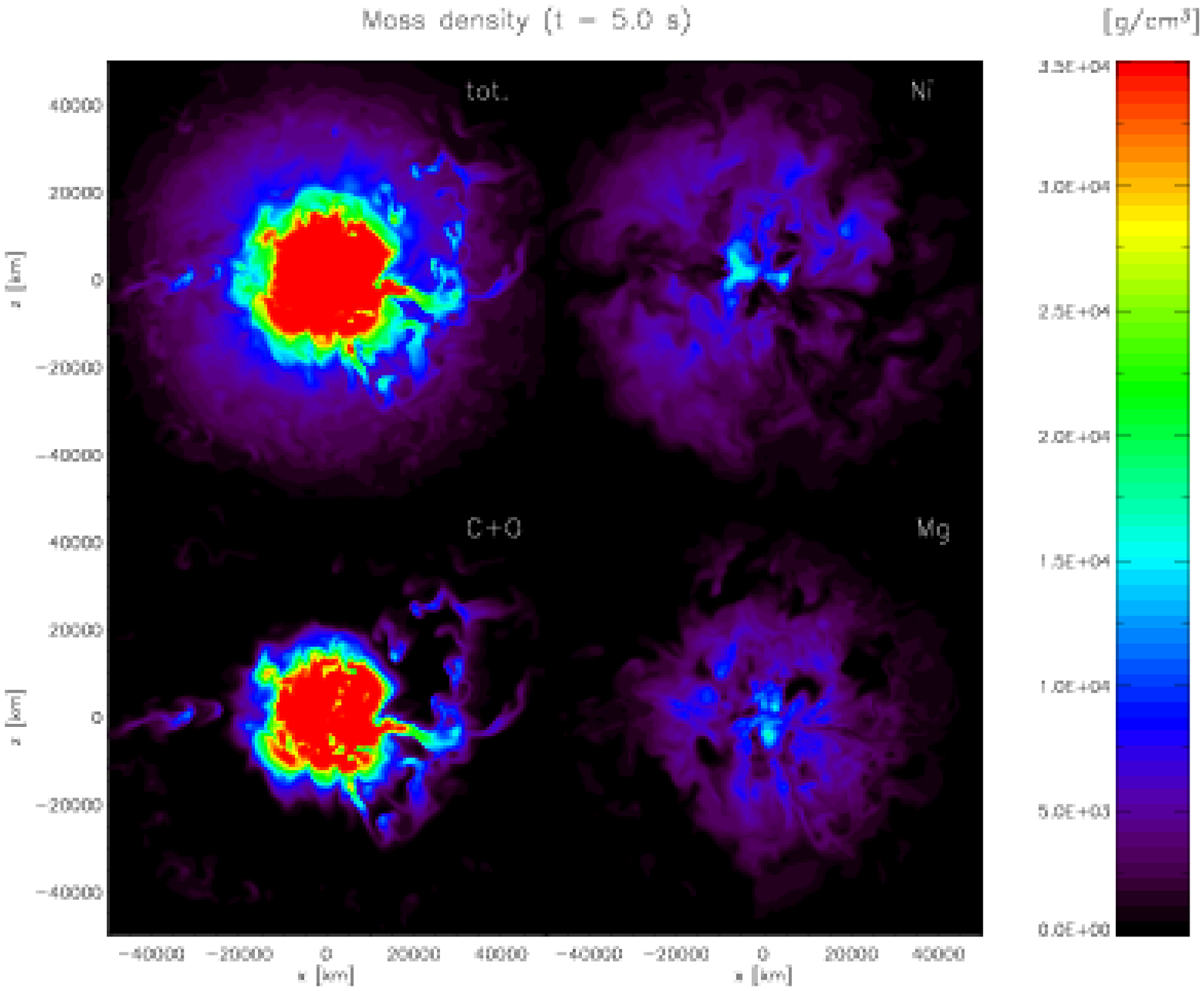}}
    \resizebox{\hsize}{!}{\includegraphics{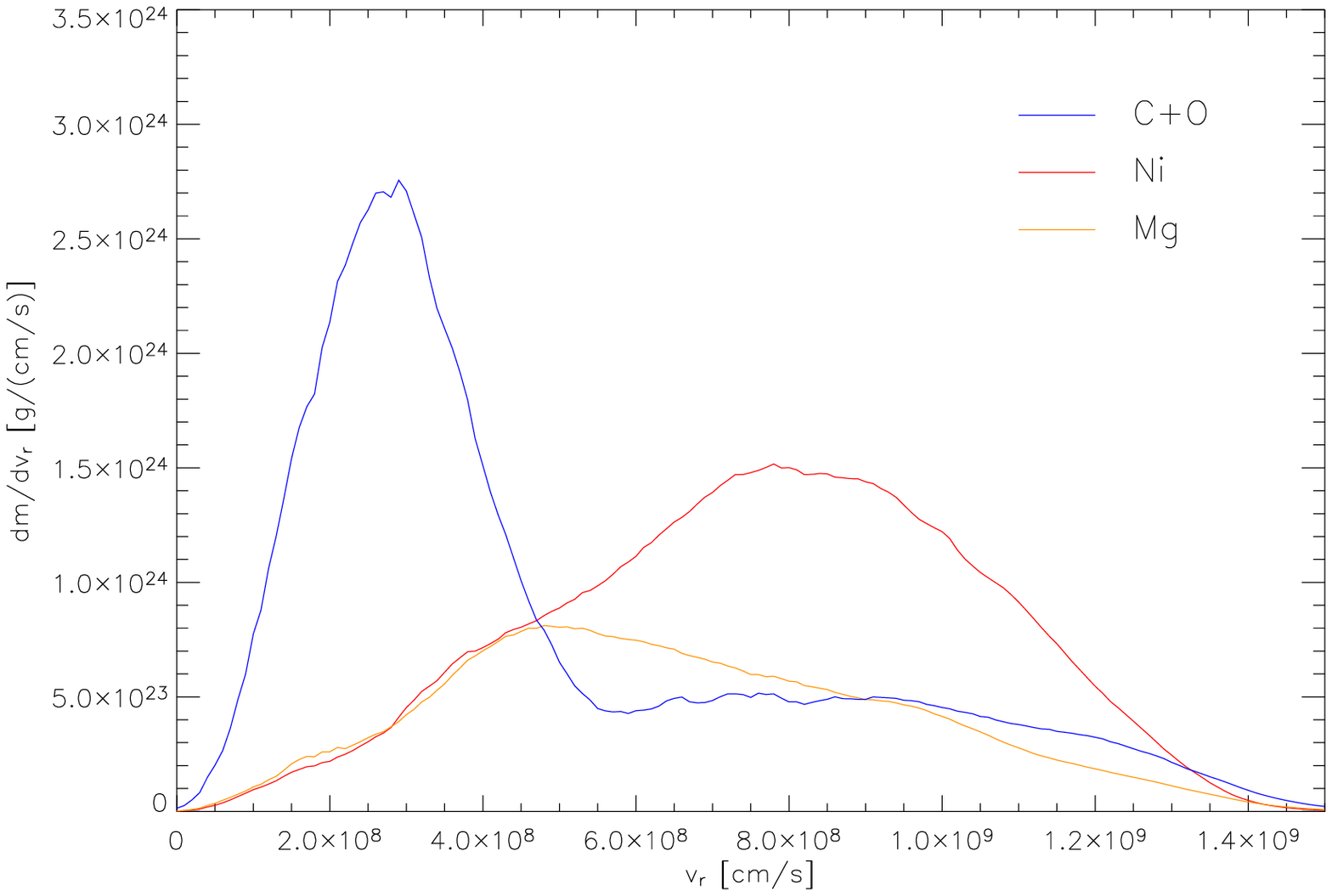}}
    \caption{Total and fractional mass densities in a central
      section and the corresponding probability density functions
      in radial velocity space for a full-star simulation with
      $C_{\mathrm{e}}=10^{2}$ at $t=5.0\,\mathrm{s}$.
    }
    \label{fg:4pi_m2}
  \end{center}
\end{figure}

\begin{figure}[thb]
  \begin{center}
    \resizebox{\hsize}{!}{\includegraphics{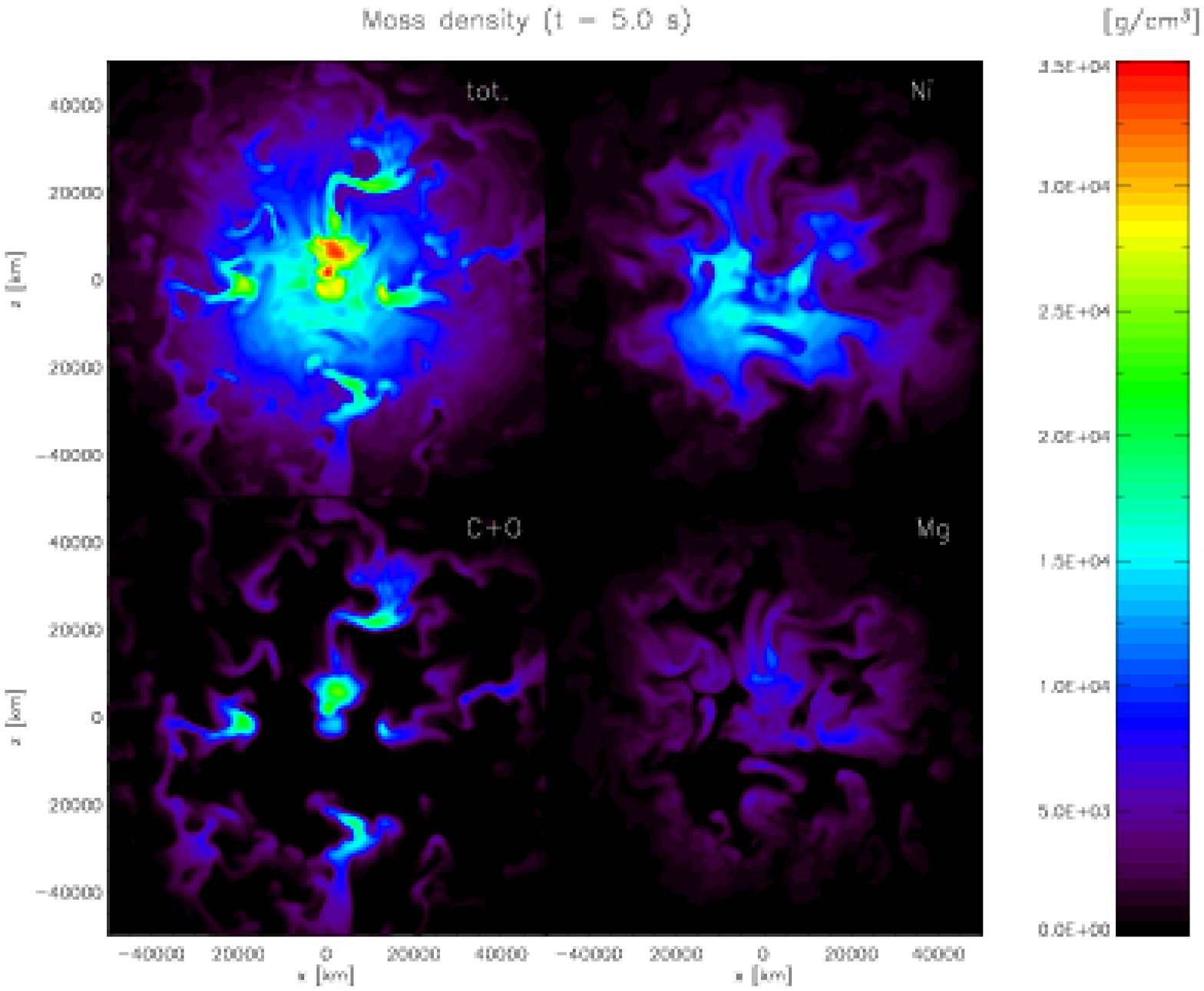}}
    \resizebox{\hsize}{!}{\includegraphics{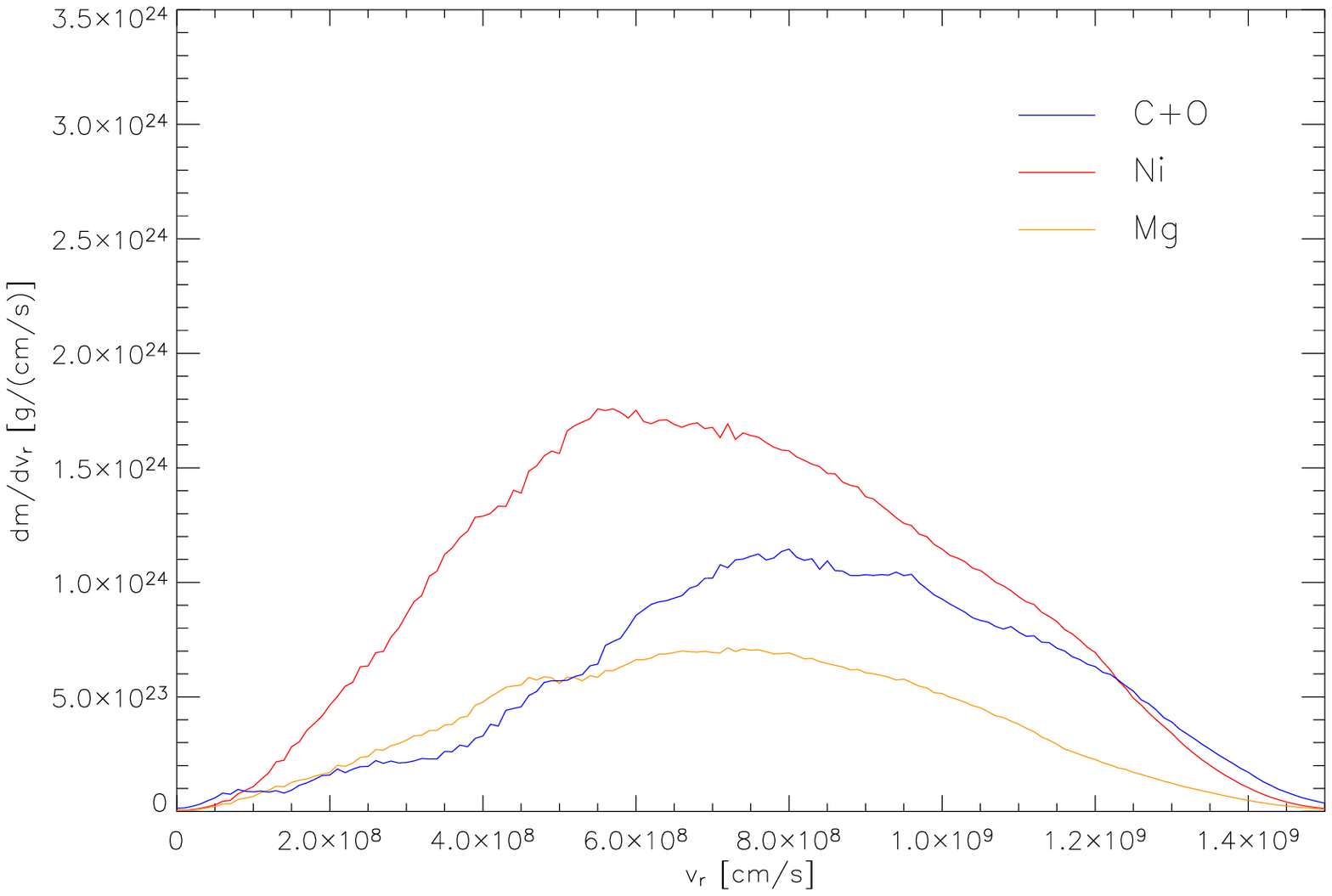}}
    \caption{Total and fractional mass densities in a central
      section and the corresponding probability density functions
      in radial velocity space for a full-star simulation with
      $C_{\mathrm{e}}=5\cdot 10^{3}$ at $t=5.0\,\mathrm{s}$.
    }
    \label{fg:4pi_m3b}
  \end{center}
\end{figure}

\begin{figure}[thb]
  \begin{center}
    \resizebox{\hsize}{!}{\includegraphics{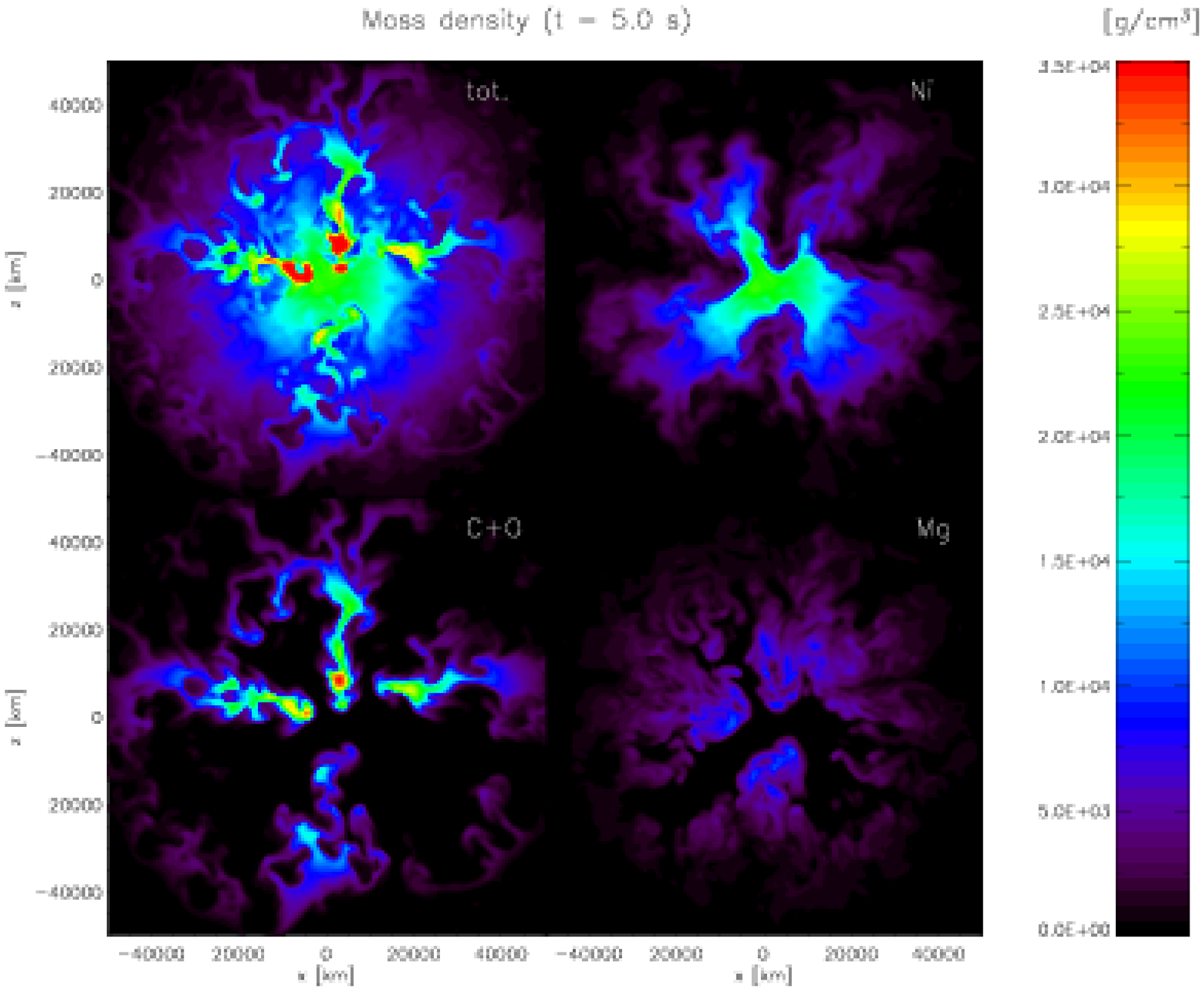}}
    \resizebox{\hsize}{!}{\includegraphics{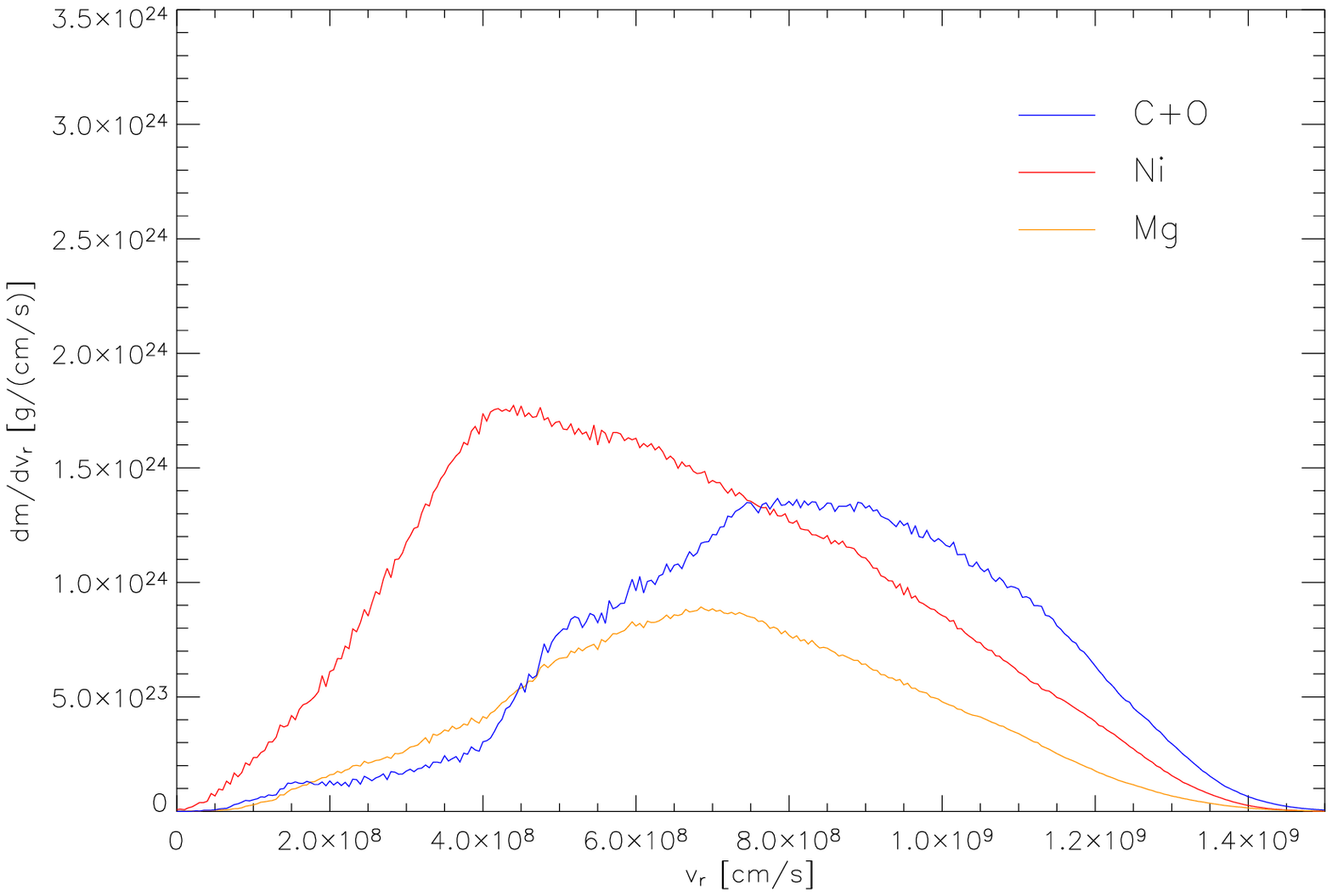}}
    \caption{Total and fractional mass densities in a central
      section and the corresponding probability density functions
      in radial velocity space for a full-star simulation with
      $C_{\mathrm{e}}=10^{5}$ at $t=5.0\,\mathrm{s}$.
    }
    \label{fg:4pi_m5}
  \end{center}
\end{figure}

Fig.~\ref{fg:4pi_m2}--\ref{fg:4pi_m5} show contour plots of the
total mass density and the partial densities of C$+$O, Mg and Ni,
respectively, in two-dimensional central sections at time
$t=5.0\,\mathrm{s}$. In comparison to Fig.~2--4 in
\citep{SchmNie06}, which show the corresponding density contours in
the simulations with the burning process terminated if $\rho_{9}\le
0.01$, one can see that the explosion ejecta are substantially
enriched in intermediate mass elements. This is also illustrated by
the plots of the fractional masses as function of radial velocity in
Fig.~\ref{fg:4pi_m2}--\ref{fg:4pi_m5}. Although the residuals of
unburned material at low radial velocities, which have been plaguing
the deflagration models, are still present in each case, it is
likely the a significant further reduction would result from volume
burning beyond the break-down of the level set description.
Nevertheless, a non-negligible amount of carbon and oxygen at radial
velocities larger than about $5000\,\mathrm{km/s}$ will be left over
for any conceivable choice of parameters.

\section{Conclusion}

We have argued that the turbulent flame propagation speed in
thermonuclear supernova simulations is given by the subgrid scale
turbulent velocity even in the distributed burning regime. Moreover,
the level set representation of the flame front remains valid beyond
the flamelet regime provided that the width of the flame brush is
smaller than grid resolution. In terms of time scales, this
constraint corresponds the termination criterion~(\ref{eq:terminate}).

Very little is known about the interaction between turbulence and the
burning process in the broken-reaction-zones regime, in which
turbulence is mixing fuel and ash faster than the nuclear reactions
are progressing. Since the break-down of the level set description
possibly occurs in this regime, we introduced the dimensionless
parameter $C^{\ast}$ which specifies the reduction of the energy
generation rate due to turbulent entrainment of fuel and ash. The
limiting case $C^{\ast}\simeq 1$ corresponds to the
thin-reaction-zones regime. In this regime, the flame is broadened due
to the enhanced diffusivity in the preheating zone while the reaction
zone is not significantly affected by turbulent eddies. As turbulence
increasingly disturbs the reaction zone, $C^{\ast}$ diminishes. For
$C^{\ast}\ll 1$, burning will be quenched.

Setting $C^{\ast}$ to values in the range from 0.01 to 1, we
performed several supernova simulations with burning being
terminated once~(\ref{eq:terminate}) was fulfilled. In these
simulations, we applied the stochastic ignition procedure described
in \citep{SchmNie06}. For $C^{\ast}\gtrsim 0.1$, a greater fraction
of intermediate mass elements was produced as in reference
simulations with termination of the burning process at the density
threshold $\rho_{9}= 0.01$. Remarkably, we found that $0.3M_{\sun}$
of intermediate mass elements was produced in the case
$C^{\ast}=1.0$ independent of the rapidity of the ignition process.
Consequently, the ignition process mainly determines the total mass of
iron group elements, whereas the production of intermediate mass
elements appears to depend solely on the progression of distributed
burning.

Although a substantial increase of the amount of iron group elements
might result from volume burning at resolved scales, which cannot be
treated within the present methodology, it seems unlikely that the
remaining fraction of carbon and oxygen would be consumed completely.
For this reason, if clear indications of carbon residuals at least in
certain type Ia supernovae were not found, delayed detonations would
be an unavoidable conclusion. Even in this case, however, studying the
physics of distributed burning in more detail might very well help to
clarify the physical mechanism of the putative DDT. On the other hand,
if burning was completed in the distributed mode in some SNe Ia,
small-scale numerical models of burning in the broken-reaction-zones
regime and explicit treatment of volume burning in large-scale
supernova simulations would be the prerequisites for further progress.

\begin{acknowledgements}
I thank Jens~C. Niemeyer and Friedrich~K. R\"{o}pke for valuable discussions.
The simulations were run on the HLRB of the Leibniz Computing Centre
in Munich. This work was supported by the Alfried Krupp Prize for
Young University Teachers of the Alfried Krupp von Bohlen und
Halbach Foundation.
\end{acknowledgements}

\bibliographystyle{aa}

\bibliography{6510}

\end{document}